\documentstyle[aps,prb,multicol,graphics]{revtex}

\begin{document}

\title{Crossover from classical to random-field critical exponents in
As-doped TbVO$_{4}$} 
\author{C.-H. Choo, H. P. Schriemer\cite{currentaddress}, and D.R.
Taylor} 
\address{Department of Physics, Queen's University, Kingston,
ON, Canada K7L 3N6} 
\date{Received \today} 
\maketitle 
\begin{abstract}
Using birefringence techniques we have measured the critical exponents 
$\beta$, $\gamma$, and $\delta$
in As-doped TbVO$_{4}$, a structural realization of the random-field
Ising model where random strain fields are introduced by V-As size
mismatch. For pure TbVO$_{4}$ we observe
the expected classical critical exponents, while for a mixed sample with
15\% As concentration our results are $\beta=0.31 \pm 0.03$,
$\gamma=1.22 \pm 0.07$ and $\delta=4.2 \pm 0.7$. These values are
consistent with the critical exponents for the short range pure Ising
model in three dimensions in agreement with a prediction by Toh. The 
susceptibility data showed a crossover with
temperature from classical to random field critical behaviour. 
\end{abstract}

\begin{multicols}{2}

Determination of the critical properties of the random field Ising
model\cite{{Belanger},{Belanger2},{Nattermann}} (RFIM) has shown encouraging progress recently,
both experimentally and theoretically, after many years of
uncertainties. Measurements on dilute antiferromagnets in a field, the
most frequently studied realization of the RFIM, have been difficult to
obtain because dilution of the magnetic species inhibits
equilibration close to the critical temperature. However recent
experiments by Slanic {\it et al.}\cite{Slanic} on samples with only 7\%
dilution have shown equilibrium behaviour through the transition
temperature, thus allowing confident determination of the critical
exponents. In the last few years, theoretical investigations making use
of a variety of analytical and computational techniques have led to
predictions of the critical exponents of the RFIM that show reasonable
consistency with each other. A recent analysis by Fortin and
Holdsworth\cite{Fortin} supports earlier suggestions that the critical
exponents for the RFIM in three dimensions ($d=3$) are those of the Ising
model for reduced dimension $d'=1.5$. Nevertheless dimensional reduction
has not been rigorously proved, and even if the effective dimension
$d'=1.5$ is correct the uncertainty in some of the calculated critical
exponents is quite large. Likewise the experimental situation is still
not satisfactory, since the critical exponents measured by Slanic {\it et
al.}\cite{Slanic} have substantial uncertainties and are only partially
consistent with theory. In addition, measurements of specific heat
critical exponents\cite{{Belanger},{Slanic2}} in dilute antiferromagnets
also appear to disagree with theory.

For another realization of the RFIM, where random strain fields are
generated by substitutional impurities in crystals undergoing structural
Ising (Jahn-Teller) transitions, the results to date have also not been
conclusive. Random fields due to As/V substitutions in DyVO$_{4}$, which
has $d=3$ Ising exponents, appeared to increase the susceptibility
exponent $\gamma$ as expected but had no effect on the order parameter
exponent $\beta$.\cite{Reza} The interpretation of the effects
of random fields in the As-doped DyVO$_{4}$ system is complicated by the
fact that the true critical behaviour of pure DyVO$_{4}$ should be
classical due to the long range strain coupling. However because of the relative
weakness of the long range to short range interactions, classical exponents
are not observable at accessible temperatures $|t| \geq 10^{-2}$, where $t=(T -
T_{\mathrm{D}})/T_{\mathrm{D}}$ is the reduced temperature,\cite{{GAGehring3},{Marques}}
leading to 
uncertainty on what the effects of the random fields would be. We have
therefore extended these experiments to the related
TbVO$_{4}$/TbAsO$_{4}$ system where the critical behaviour of the pure
compounds is unequivocally
classical\cite{{Harley},{Elliot},{Wells},{Sandercock},{Berkhan}}, and
searched for changes in critical behaviour in mixed crystals due to
random fields. Since this system starts from a different universality
class, the results cannot be compared directly with results from dilute
antiferromagnets, but it is an important system that can
independently test theoretical models and predictions of random field
effects. In contrast with the large number of theoretical investigations
of the short-range RFIM, the literature on the random-strain
version of the RFIM is very limited, consisting primarily of a paper by
Toh.\cite{Toh} In this paper, Toh compares the random-strain RFIM with
long range forces to that of the short-range RFIM under a
renormalization group analysis. The main result is that the critical
exponents should change from classical values to values that are close
to those of the $d=3$ pure short-range Ising model. 

The lowest $4f$ electronic levels in TbVO$_{4}$ consists of 2 singlet
states $\sim 18$ cm$^{-1}$ apart and a non-Kramers' doublet
approximately halfway between. Coupling between the doublets and lattice
distortions leads to a Tb ion-ion interaction of the Ising
form\cite{Elliot} and a tetragonal-orthorhombic phase transition at
temperature T$_{\mathrm{D}}$. Since the Tb ions are coupled
predominantly to $k\sim 0$ acoustic phonons and to bulk strains, the
ion-ion interaction is very long range. The order parameter is the
macroscopic strain $a-b$ where $a$ and $b$ are basal plane unit cell
parameters in the orthorhombic phase. The orthorhombic distortion gives
rise to birefringence, $\Delta n$, which is proportional to $a-b$ (at
least to a good approximation\cite{Harley}).

The full Hamiltonian for the coupled electron-phonon TbVO$_{4}$ system
in an external magnetic field can be written as
\begin{equation}
H=- \case1/2 \sum_{ij} J_{\it ij}\sigma^{\it z}_{\it i} \sigma^{\it z}_{\it j} -
\case1/2 \epsilon \sum_{\it i} (1 + \tau^{\it z}_{\it i})\sigma^{\it x}_{\it i} -
\text{B}\sum_{\it i} {\it m^{x}_{i}} 
\label{meanfield} \end{equation} 
where $m^{x}_{i} = \case1/4{\text{g}}\mu_{\text{B}}(1 + \sigma^z_{i})\tau^x_{i}$ 
and $\sigma^{z}, \sigma^{x}, \tau^{z}
\text{and } \tau^{x}$ are Pauli type
operators\cite{{Elliot},{Gehring2}}. $J_{ij}$ describes the ion-ion
interactions, $2\epsilon$ is the high temperature splitting between the
outer singlets and B is the magnetic field applied along the $x$ (or
$a$) axis (i.e. along the 110 direction). The field B is able to induce an
orthorhombic distortion because of the strongly anisotropic Tb magnetic
moment in the orthorhombic phase. In the mean field approximation, a
Landau expansion of the free energy, with $\epsilon=0$, shows that
B$^{2}$/T is effectively an ordering field,\cite{Page} and this 
is supported by the experimental data
that follows. For B and $\epsilon$ small, TbVO$_{4}$ is well described
by an Ising model Hamiltonian. As the mode softening at the transition
in this type of system is anisotropic\cite{Cowley}, classical critical
exponents are expected, and observed, rather than $d=3$ Ising exponents. 

In the mixed compound, Tb(As$_{x}$V$_{1-x}$)O$_{4}$, a fraction $x$ of
the V atoms are replaced by As atoms, generating random, static
strain fields, one component of which has the right symmetry to couple
to the order parameter.\cite{GAGehring2} For $\epsilon=\text{B}=0$ the
Hamiltonian has the form of the RFIM,
\begin{equation} H =-\case1/2
\sum_{ij} J_{\it ij} \sigma^{\it z}_{\it i} \sigma^{\it z}_{\it j} -
\sum_{\it i}h_{\it i}\sigma^z_{\it i} \label{randomfield} 
\end{equation} 
where $h_{i}$ is
the random local strain field which is expected to have a Gaussian distribution about
$h=0$. In his analysis of this type of random field system with
anistropic mode softening, Toh predicts changes to the upper critical
dimension, thus modifying the values of critical exponents at $d=3$.

Crystals of Tb(As$_{x}$V$_{1-x}$)O$_{4}$ with impurity concentrations of
$x= 0$ and 0.15 were prepared using the flux growth method at the
University of Oxford. The crystals were cut and
polished perpendicular to the $c$ axis, with thickness of about 1 mm. They
were mounted in a strain-free manner with the $c$ axis horizontal in a
helium optical cryostat. The crystal could be rotated about a vertical
axis, allowing alignment of the $c$ axis parallel to the laser beam and
one of the $a$ axes parallel to a horizontal magnetic field. A circular 
aperture in the sample holder of about 1 mm in diameter limited the
sampled area to a small region of the crystal, and thus 
reduced effects due to any inhomogeneous composition, temperature and
ordering field
that may be present in the crystal.

The light source for birefringence experiments was a HeNe laser 
operating at 543.5
nm, a wavelength that should give reasonable birefringence in this
crystal.\cite{Hikel} Photoelastic modulation and lock-in detection 
allowed sensitive measurement\cite{Ferre} of the phase shift $\phi$ due 
to the orthorhombic distortion. Adjustment of a Babinet compensator 
ensured that the detector output was proportional 
to $\phi$. In a typical experiment, the data acquisition system brought the sample 
to each desired temperature, waited up to 15 minutes for equilibration, and then 
ramped the magnetic field up and down while recording field, light intensity, 
and temperature.

Below T$_{\mathrm{D}}$, 
the orthorhombic distortions are equally likely
to be in the 110 or 1$\overline{1}$0 direction, leading to the formation
of twinned orthorhombic domains separated by domain walls. To avoid
the cancellation of birefringence due to 
multiple domains, the crystal is forced into a
single domain by applying a magnetic field that favours the distortion
axis parallel to the magnetic field.\cite{KAGehring,GAGehring1} Thus to
determine the birefringence, and hence $\beta$, below T$_{\mathrm{D}}$,
we recorded data over a range of magnetic fields and extrapolated the
high magnetic field data back to zero field. With the rather small
magnetic fields available ($< 0.35$ T) it was sometimes difficult to
achieve a single domain at temperatures close to T$_{\mathrm{D}}$ , at
$|$T - T$_{\mathrm{D}}| < 0.2 $ K for $x = 0.15$, but this became
progressively easier further away from T$_{\mathrm{D}}$. For the $x=0$
 sample where pinning is presumably weaker, single domain structure at
$|$T - T$_{\mathrm{D}}| < 0.75 $ K was relatively easy to achieve. For
this reason, the critical isotherm exponent, $\delta$, obtained from the
dependence of the induced birefringence on ordering field at the
transition temperature, is more reliable for pure TbVO$_{4}$ than that
for the mixed sample. Above T$_{\mathrm{D}}$, the change in induced
birefringence resulting from a change in the ordering field gives the
susceptibility and hence $\gamma$. Although the exponents for pure
TbVO$_{4}$ are known to be classical,\cite{Harley}, their measurement 
provides a comparison with the
mixed sample where modified critical exponents are expected.
In experiments, T$_{\mathrm{D}}$ appeared to vary slightly from run to run  
because of effects such as mounting strains and temperature gradients. 
To minimize the effects of a variable T$_{\mathrm{D}}$ on the results, 
efforts were made to measure
all exponents of one particular sample on the same run. In the power-law 
fits to the data, T$_{\mathrm{D}}$ was chosen to optimize both the $\beta$ and
$\gamma$ fits simultaneously. The consistency in the results of repeated
experiments and the quality of the fits give confidence in the
results.

\begin{figure}
\vspace*{4.0cm}
\begin{center}
\scalebox{0.75}{\includegraphics[0,0][7.9in,8.7in]{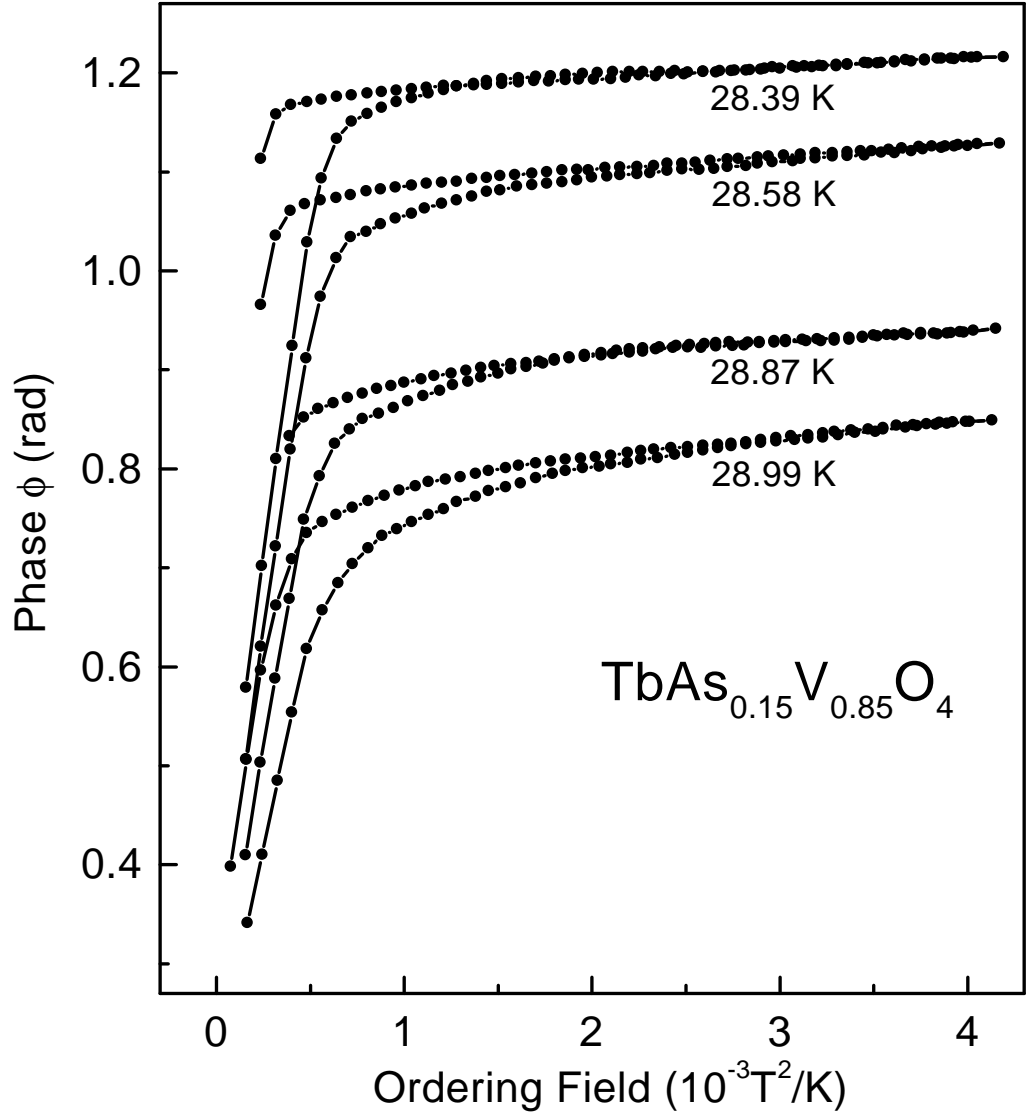}}
\vspace{0.2cm}
\caption{Changes in phase $\phi$ (birefringence) in $x=0.15$ sample at selected 
temperatures below T$_{\mathrm{D}}$ for increasing (lower points) and decreasing 
(upper points) ordering fields. The lines join data points.} 
\label{figure1}
\end{center}
\end{figure}

\begin{figure}
\vspace*{4.0cm}
\begin{center}
\scalebox{0.75}{\includegraphics[0,0][7.5in,9.0in]{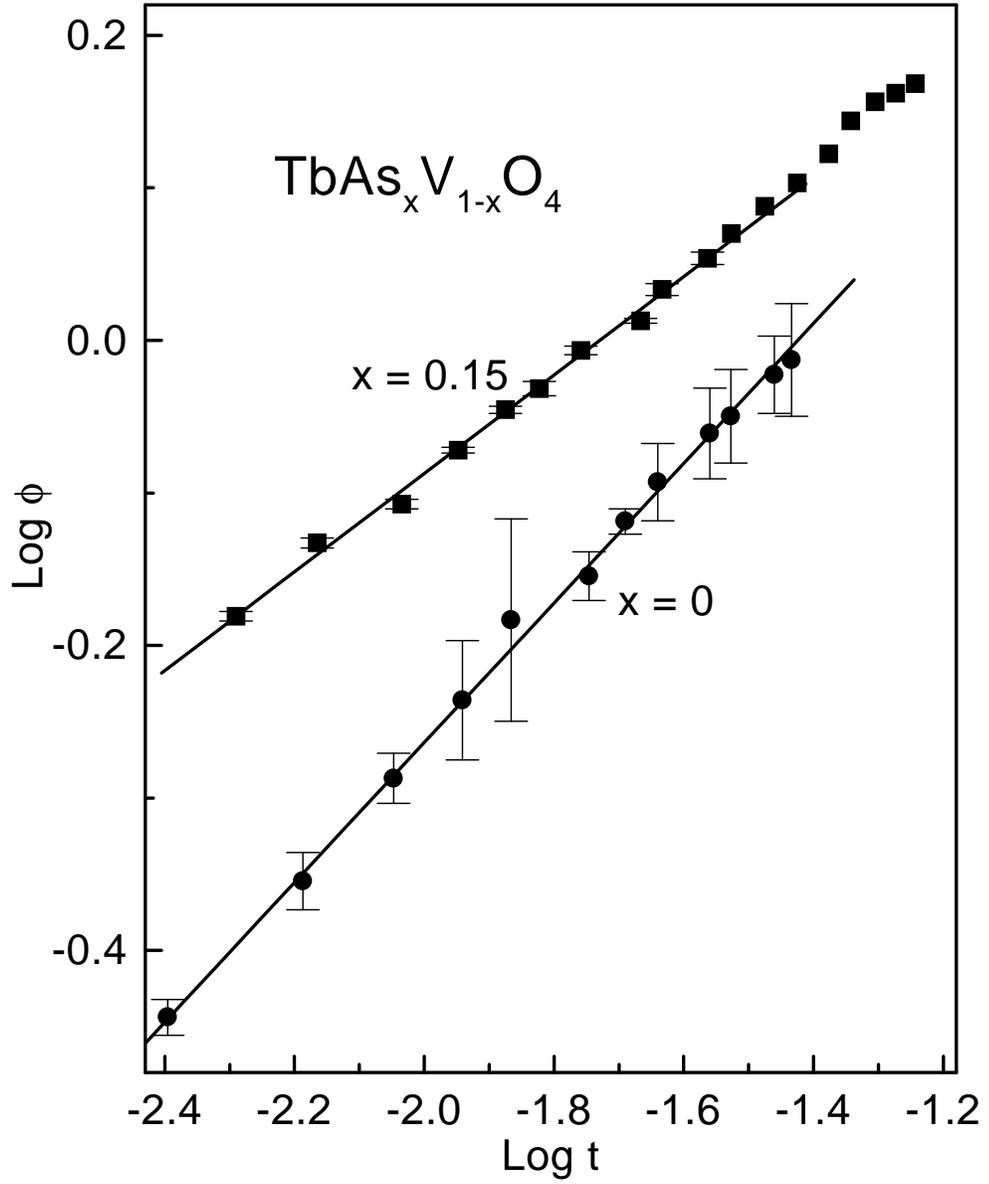}}
\vspace{0.2cm}
\caption{Log-log plots of birefringence {\it versus} reduced temperature
for $x=0$ and $0.15$. The slopes give the order parameter critical exponent $\beta$.} 
\label{figure2}
\end{center}
\end{figure}

The birefringence with increasing and decreasing ordering fields for
various temperatures below T$_{\mathrm{D}}$ in the $x=0.15$ sample is
shown in Fig.\ \ref{figure1}. Hysteresis observed in the low field
region is attributed to pinning of the multidomain structure.
The data from the higher field region where hysteresis is absent were
extrapolated to determine the zero field birefringence. A log-log plot
of both the $x=0.15$ and $x=0$ data, fitted to a power law of the form
$\Delta n \propto |t|^{\beta}$ is shown in Fig.\ \ref{figure2}. Using
T$_{\mathrm{D}} = 29.26 \pm 0.03 $ K and data in the range $0.005 < |t|
<0.027$, we obtained a value of $\beta = 0.31 \pm 0.03$ for the mixed
sample. At larger $|t|$ we found no convincing evidence for crossover 
behaviour towards the classical exponent. For the pure sample, 
we obtained $\beta = 0.46 \pm 0.06$ for T$_{\mathrm{D}} = 32.32 \pm 0.04 $ K 
from data in the range $0.004 < |t| <0.037$.

\begin{figure}
\vspace*{4.0cm}
\begin{center}
\scalebox{0.75}{\includegraphics[0,0][7.6in,9.1in]{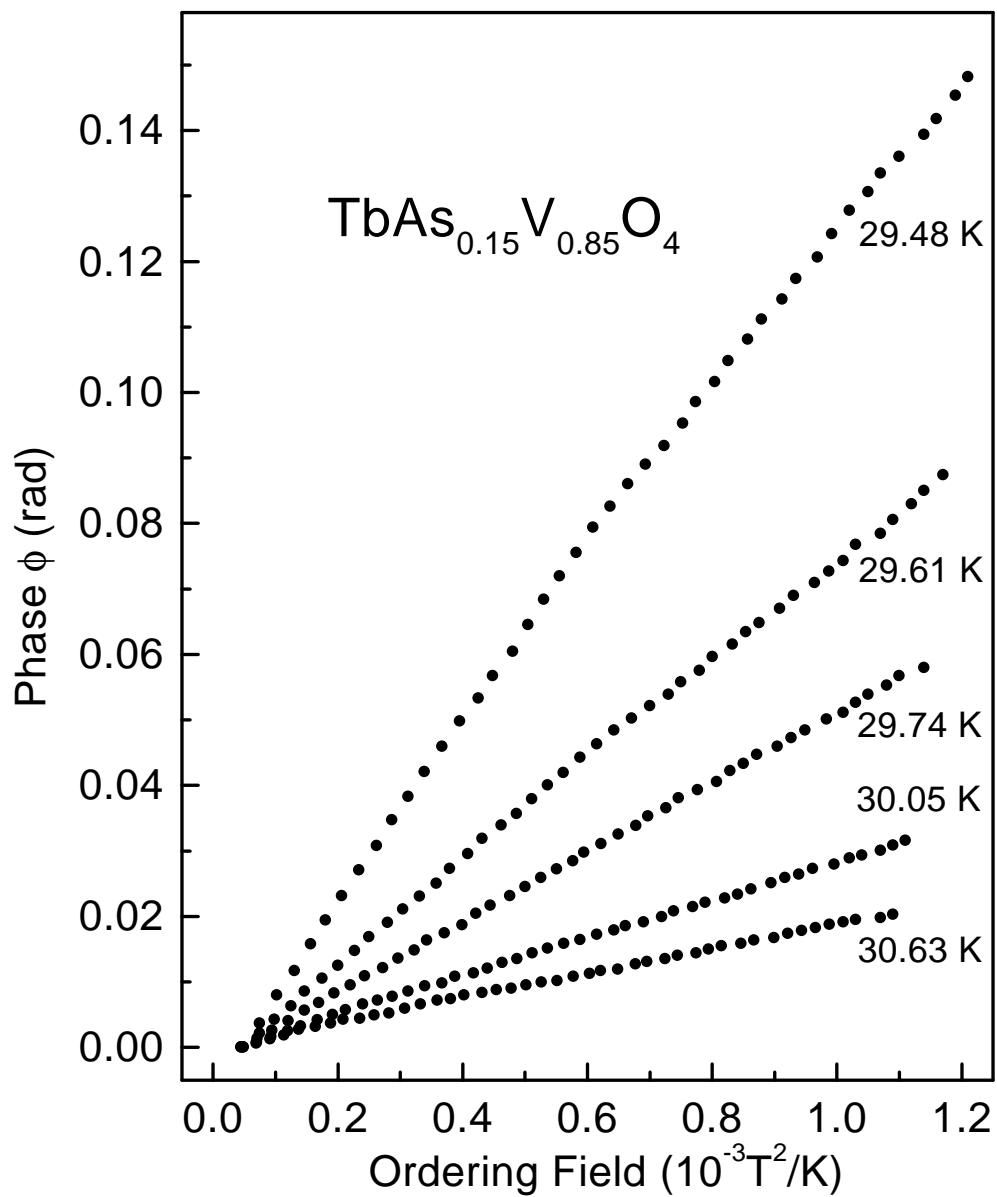}}
\vspace{0.2cm}
\caption{Changes in birefringence {\it versus} ordering field in $x=0.15$ sample
at selected temperatures above T$_{\mathrm{D}}$. Slopes in the small-field limit give 
susceptibilities.} 
\label{figure3}
\end{center}
\end{figure}

\begin{figure}
\vspace*{4.0cm}
\begin{center}
\scalebox{0.75}{\includegraphics[0,0][8.1in,9.0in]{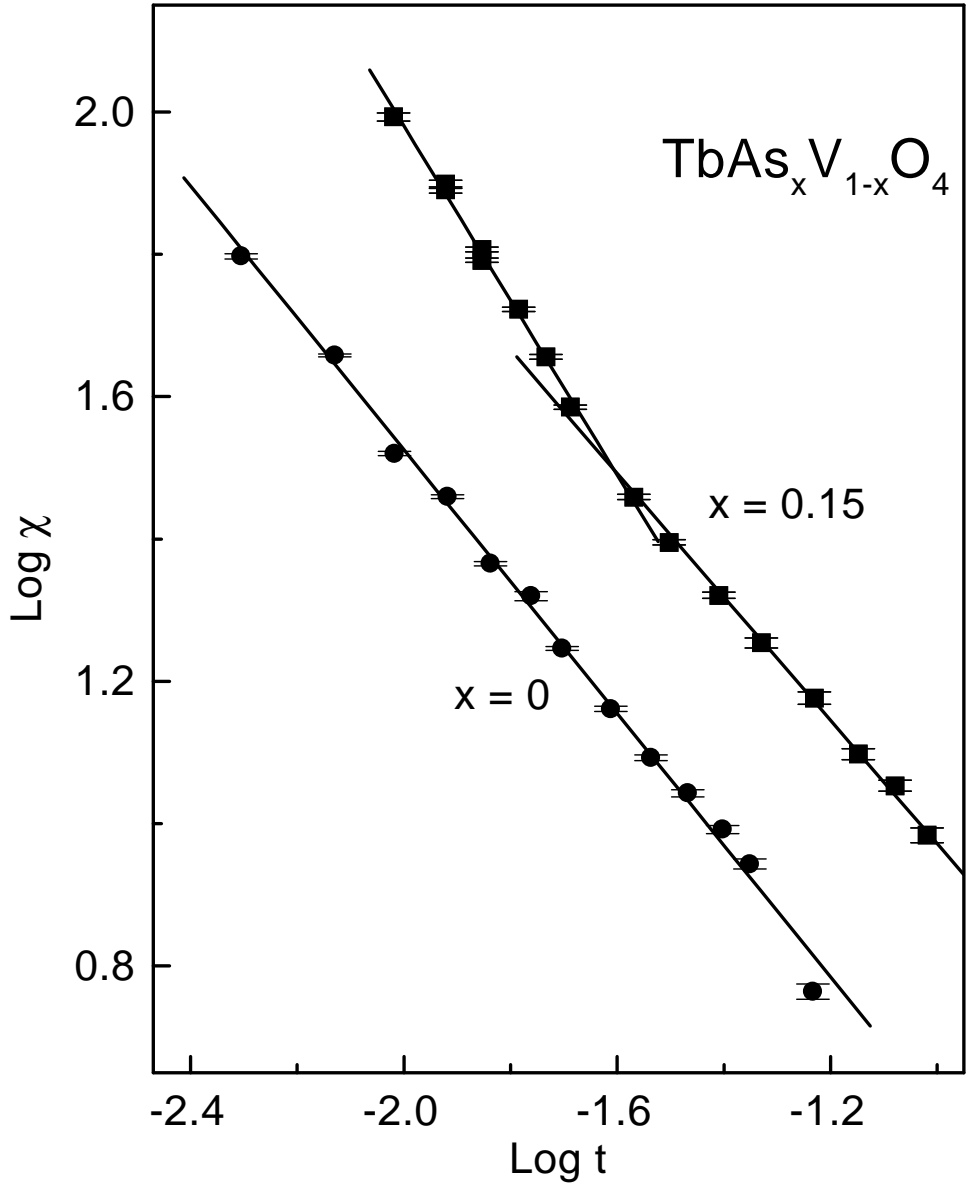}}
\vspace{0.2cm}
\caption{Log-log plots of susceptibility {\it versus} reduced temperature for
$x=0$ and $0.15$. The data for the mixed sample show a crossover from classical to 
random-field critical behaviour.} 
\label{figure4}
\end{center}
\end{figure}

Figure\ \ref{figure3} presents susceptibility data for selected 
temperatures above T$_{\mathrm{D}}$. At temperatures close to
T$_{\mathrm{D}}$, these slopes deviate from linearity, attributable 
in part to the presence of non-zero birefringence at
zero field, probably caused by internal strains\cite{Harley}. The
susceptibility exponent $\gamma$ was found by fitting the various
values of $\chi(T)$ to the relation $\chi^{-1} \propto |t|^{\gamma}$.
Figure\ \ref{figure4} shows a log-log plot of the susceptibility $\chi$ {\it versus} 
the reduced temperature $|t|$ for $x =0$ and $0.15$ samples. As can be
seen from Fig.\ \ref{figure4}, the log-log plot for $x=0.15$ shows two
linear fits with the data closer to T$_{\mathrm{D}}$ having a larger
slope than that for data further away. Data in the range $0.027 > |t| > 0.01$ 
(closest to T$_{\mathrm{D}}$) were optimized first
to a linear fit giving T$_{\mathrm{D}} = 29.26 \pm 0.03 $ K and 
$\gamma = 1.22 \pm 0.07$. A linear
fit to data in the range $0.100 > |t| > 0.021$ (further from T$_{\mathrm{D}}$) 
with T$_{\mathrm{D}}$ unchanged yields a value of $\gamma =
0.89 \pm 0.03$. We did not attempt to fit these data to a crossover 
function, but they suggest a crossover temperature near $|t|=0.024$ 
($29.96$ K). The log-log plot of the susceptibility {\it versus} reduced temperature for pure
TbVO$_{4}$ in Fig.\ \ref{figure4} does not show a crossover effect. A power 
law fit in the range $0.058 > |t| > 0.005$ gives T$_{\mathrm{D}} =
32.32 \pm 0.04 $ K and $\gamma = 0.92 \pm 0.07$.

Nonlinearity in the dependence of the order parameter on the ordering
field at temperatures below T$_{\mathrm{D}}$ can be attributed to the
progressive detwinning of the sample with increasing field\cite{Reza2}
(i. e. changing from a multidomain to a single domain structure). The
critical isotherm exponents $\delta$ were extracted from the field-induced 
birefringence data at temperatures closest to that of the
previously determined transition temperatures for $x=0.15$ and $x=0$
samples. Only the higher field $\Delta$n data were fitted to the power
law (B$^{2}$/T$_{\mathrm{D}})^{1/\delta}$. For the $x=0.15$ sample,
we obtained a value of $\delta = 4.2 \pm 0.7$ at $T=29.24 \pm 0.03$ K
and for the pure sample $\delta = 2.6 \pm 0.4$ at $T=32.33 \pm 0.03$ K. 
The uncertainty in locating T$_{\mathrm{D}}$ is 
incorporated in the uncertainty in $\delta$.

Our values of the critical exponents for pure TbVO$_{4}$ agree
satisfactorily with the values 
$\beta=0.5, \gamma=1$, and $\delta=3$ expected for a mean field system. 
For the random-field sample our values $\beta=0.31
\pm 0.03$, $\gamma=1.22 \pm 0.07$, and $\delta = 4.2 \pm 0.7$ are in 
good agreement with the exponents for the $d=3$
Ising model,\cite{Guillou} $\beta=0.33$, $\gamma=1.24$, and $\delta=4.8$. 

Toh's predictions\cite{Toh} for the effects of random fields on the exponents 
in this type of system are therefore well supported by our results. Toh noted 
that while the
anisotropic strain interactions in TbVO$_{4}$ reduce the upper critical
dimension $d^{*}$ from 4 for the standard Ising model to 2, resulting in
classical exponents\cite{Cowley}, the introduction of
random strain fields raises $d^{*}$ from 2 to 4 again, leaving the
critical exponents the same as for the standard Ising model in $d=3$. 
In view of the focus in recent theoretical analysis on dimensional
reduction it is of interest to examine Toh's analysis and our results in
terms of changes in effective dimension $d'$ instead of in $d^{*}$. Thus
for the Ising model we regard the upper critical dimension to be fixed
at 4; in pure TbVO$_{4}$, the anisotropic strain interactions presumably
raise the effective dimension from 3 to 5, giving classical exponents as
before. If changes in critical properties due to random fields can be 
described by dimensional reduction, the consensus prediction\cite{{Nattermann},{Fortin}} 
is that random fields reduce the effective dimensionality by $2-\eta$, 
where $\eta$ is the exponent for the decay of magnetization correlations. 
In the present case, since $\eta=0.032$ for $d=3$,\cite{Guillou} the
dimensionality should be reduced by 1.97 or essentially 2, resulting in $d'=3$. 
Hence our results, within their accuracy, are consistent
with dimensionality reduction by random fields of $2-\eta \sim 2$ for this system.

\end{multicols}

\clearpage

\clearpage

\clearpage

\clearpage

\end{document}